

\input mn
\input epsf


\let\sec=\section
\let\ssec=\subsection


\font\japit = cmti10 at 11truept
\def\ss{\scriptscriptstyle\rm}
\def\ref{\parskip =0pt\par\noindent\hangindent\parindent
    \parskip =2ex plus .5ex minus .1ex}
\def\gs{\mathrel{\lower0.6ex\hbox{$\buildrel {\textstyle >}
 \over {\scriptstyle \sim}$}}}
\def\ls{\mathrel{\lower0.6ex\hbox{$\buildrel {\textstyle <}
 \over {\scriptstyle \sim}$}}}
\newcount\equationo
\equationo = 0

\newcount\fred
\fred=0

\def\outeqn#1{\the #1}

\def\leftdisplay#1$${\leftline{$\displaystyle{#1}$
  \global\advance\equationo by1\hfill (\the\equationo )}$$}
\everydisplay{\leftdisplay}

\def\erf{{\rm erf}\,}

\def\kms{\;{\rm km\,s^{-1}}}

\def\hompc{\,h\,{\rm Mpc}^{-1}}
\def\mpcoh{\,h^{-1}\,{\rm Mpc}}

\def\japfig#1#2#3#4{
\ifnum #2 = 1 
\beginfigure{#1}
\epsfxsize=8.2cm
\centerline{\epsfbox[60 208 510 588]{#3}}
\fi
\ifnum #2 = 2 
\beginfigure{#1}
\epsfxsize=8.2cm
\centerline{\epsfbox[53 15 465 785]{#3}}
\fi
\ifnum #2 = 3
\beginfigure*{#1}
\epsfxsize=15.5cm
\centerline{\epsfbox{#3}}
\fi
\ifnum #2 = 4
\beginfigure{#1}
\epsfxsize=8.2cm
\centerline{\epsfbox{#3}}
\fi
\caption{%
{\bf Figure #1.}
#4
}
\endfigure
}



\def\aj{AJ}

\def\apj{ApJ}

\def\mn{MNRAS}

%

\pageoffset{-0.8cm}{0.2cm}




\begintopmatter  

\vglue-2.2truecm
\centerline{\japit Accepted for publication in Monthly Notices of the R.A.S.}
\vglue 1.7truecm

\title{Halo occupation numbers and galaxy bias}

\author{J.A. Peacock \& R.E. Smith}

\affiliation{
Institute for Astronomy, University of Edinburgh, 
Royal Observatory, Blackford Hill, Edinburgh EH9 3HJ}

\shortauthor{J.A. Peacock \& R.E. Smith}

\shorttitle{Halo occupation numbers and galaxy bias}


\abstract{%
We propose a heuristic model that displays
the main features of realistic theories for galaxy bias.
We first show that the low-order clustering statistics 
of the dark-matter distribution depend almost entirely
on the locations and density profiles of dark-matter
haloes. The quasilinear mass correlations are in fact reproduced
well by a model of independent randomly-placed haloes.

The distribution of galaxies within the halo density field depends on (i) the
efficiency of galaxy formation, as manifested by the
{\it halo occupation number} -- the number of galaxies
brighter than some sample limit contained in a halo of a given
mass; (ii) the location of these galaxies within their halo.
The first factor is constrained by the empirical luminosity
function of groups. For the second factor, we assume that
one galaxy marks the halo centre, with any remaining
galaxies acting as satellites that trace the halo mass.
This second assumption is essential if
small-scale galaxy correlations are to remain close to a single power law,
rather than flattening in the same way as the correlations
of the overall density field.

These simple assumptions amount to a recipe for
non-local bias, in which the probability of finding
a galaxy is not a simple function of its local mass density.
We have applied this prescription to some CDM models of current interest,
and find that the predictions are close to the observed galaxy correlations 
for a flat $\Omega=0.3$ model ($\Lambda$CDM), but not for
an $\Omega=1$ model with the same power spectrum ($\tau$CDM).
This is an inevitable consequence of cluster normalization for the
power spectra: cluster-scale haloes of given mass have smaller
core radii for high $\Omega$, and hence display enhanced small-scale clustering.
Finally, the pairwise velocity dispersion of galaxies in the $\Lambda$CDM
model is lower than that of the mass, allowing cluster-normalized
models to yield a realistic Mach number for the peculiar
velocity field. This is largely due to the strong variation of
galaxy-formation efficiency with halo mass that is required in this model.
}


\maketitle  

\sec{INTRODUCTION}

The large-scale structure in the distribution of galaxies has 
long been assumed to arise from primordial inhomogeneities
in the cosmological mass distribution. The quantitative
study of the evolution of these inhomogeneities is now
a mature field, particularly in the case of universes
dominated by collisionless cold dark matter. Large
$N$-body simulations have established the clustering
properties of the CDM density field, and shown how they can
be understood in terms of simple nonlinear scaling
arguments (e.g. Jenkins et al. 1998).

The outstanding challenge is of course the connection with
the galaxy distribution. The Poisson clustering hypothesis
would propose that galaxies are simply a dilute sampling
of the mass field. If this were a correct hypothesis,
no CDM universe would be acceptable, since the correlation
functions for these models differ from the observed galaxy
correlations in a complicated scale-dependent fashion
(e.g. Klypin, Primack \& Holtzman 1996;
Peacock 1997; Jenkins et al. 1998). Allowing the galaxy
density to be any local function of the mass density
does not remove this problem (Coles 1993; Mann, Peacock \& Heavens 1998).
It may well be that CDM models are not a good description of reality,
but the problems with correlation functions are not a very
strong argument in this direction, for the principal reason
that the formation of galaxies must be a non-local process to
some extent. The modern paradigm was introduced by White \& Rees (1978):
galaxies form through the cooling of baryonic material in
virialized haloes of dark matter. The virial radii of these
systems are in excess of 0.1~Mpc, so there is the potential
for large differences in the correlation properties of
galaxies and dark matter on these scales.

A number of studies have indicated that the observed
galaxy correlations may indeed be reproduced by CDM models.
The most direct approach is a numerical simulation that
includes gas, and relevant dissipative processes.
This is challenging, but just starting to be feasible with
current computing power (Pearce et al. 1999). The alternative
is `semianalytic' modelling, in which the merging history
of dark-matter haloes is treated via the extended Press-Schechter
theory (Bond et al. 1991), and the location of galaxies within
haloes is estimated using dynamical-friction arguments
(e.g. Kauffmann et al. 1993, 1999; Cole et al. 1994; 
Somerville \& Primack 1999;
van Kampen, Jimenez \& Peacock 1999;
Benson et al. 2000a,b). Both these approaches
have yielded similar conclusions, and shown how CDM models
can match the galaxy data: specifically, the low-density
flat $\Lambda$CDM model that is favoured on other
grounds can yield a correlation function that is close to a
single power law over $1000 \gs \xi \gs 1$, even though the
mass correlations show a marked curvature over this range
(Pearce et al. 1999; Benson et al. 2000a).

These results are impressive, yet it is frustrating to have a conclusion
of such fundamental importance emerge from a complicated
calculational apparatus. The aim of this paper is therefore to
isolate the main processes that produce the effect, yielding
a simple model for the galaxy distribution that results from
a given density field. Such a model is not a substitute for the
full physical calculations, but it should have pedagogical
value, and also be of practical use in setting up large
simulated galaxy surveys, as well as investigating what range
of CDM models can be made consistent with observation.
We shall argue that the main features of the galaxy density
field can be understood in terms of a model where the key
feature is the halo occupation number: the number of galaxies
found above some luminosity threshold in a virialized halo of a given mass.
To some extent, this is a very old idea, going back at least
to Neyman, Scott \& Shane (1953). More recent manifestations
have emphasized that nonlinear mass correlations are closely
related to the density profiles of virialized dark-matter haloes
(Sheth \& Jain 1997; Yano \& Gouda 1999; Ma \& Fry 2000), and that the
clustering and dynamical properties of galaxies may be affected
by an efficiency of galaxy formation that depends on halo mass
(Jing, Mo \& B\"orner 1998; Seljak 2000). 
The main new features introduced in the present paper are to apply these
arguments including significant recent revisions to our ideas
about halo density profiles (Moore et al. 1999; M99), and especially
to argue that the required halo occupation numbers can
be constrained by other data (principally the group luminosity function).
This removes the arbitrary degree of freedom corresponding
to the mass-dependence of the efficiency of galaxy formation, and allows
relatively robust model predictions.

The structure of the paper is as follows. Section 2 summarizes
models for the nonlinear density correlations, and shows that
these can be understood quantitatively in terms of the
density profiles of virialized haloes. Section 3 investigates
how the correlations are affected by (i) the occupation
number (the number of galaxies per halo, and how this
varies with mass); (ii) the placement of galaxies within haloes.
The first factor appears to control the bias on intermediate
scales; the second determines the small-scale correlations.
We argue that these degrees of freedom are already constrained 
by empirical data on galaxy groups.
Section 4 then looks at detailed numerical properties of the galaxy
field in this approximation. The galaxy power spectrum is
reproduced quite robustly in shape and amplitude, independent
of the cosmological model, except on the largest scales.
It is also possible to understand the long-standing problem of the low Mach number
of the observed galaxy distribution. Finally, Section 5 sums up.

\strut

\bigskip

\section{CORRELATIONS OF INDEPENDENT HALOES}

\ssec{Correlation functions}

One of the earliest suggested models for the
galaxy correlation function was to consider a density
field composed of randomly-placed independent clumps
with some universal density profile (Neyman, Scott \& Shane 1953; Peebles 1974).
Since the clumps are placed at random (with number density $n$), the only excess neighbours
to a given mass point arise from points in the same clump, and the correlation 
function is straightforward to compute in principle; see Appendix A for details.
For the case where the clumps have a power-law density profile,
$$
\rho= n B r^{-\epsilon},
$$
truncated at $r=R$, the small-$r$ behaviour
of the correlation function is $\xi\propto r^{3-2\epsilon}$,
provided $3/2 < \epsilon <3$. For smaller values of $\epsilon$, $\xi(r)$
tends to a constant as $r\rightarrow 0$.
In the isothermal $\epsilon=2$ case, the correlation function for
$r\ll R $ is
$$
\xi(r)={\pi^2 B\over 4 rR} = {\pi N \over 16 r R^2 n},
$$
where $N$ is the total number of particles per clump
(Peebles 1974).

The general result is that the correlation function is
less steep at small $r$ than the clump density profile, which
is inevitable because an autocorrelation function involves convolving
the density field with itself.
A long-standing problem for this model is therefore that the
predicted correlation function is much flatter than is
observed for galaxies: $\xi \propto r^{-1.8}$ is the
canonical slope, apparently requiring 
clumps with very steep density profiles, $\epsilon=2.4$.

Despite this difficulty, we will argue below that this model is
capable of giving a good understanding of the properties of the cosmological
density field. Two small alterations are required to the
discussion so far, replacing the arbitrary power-law clumps
with the realistic density profiles of virialized
dark-matter haloes, and allowing for a dispersion in halo masses.
The properties of dark-matter haloes have been well
studied in $N$-body simulations, and highly accurate
fitting formulae exist, both for the mass function and
for the density profiles. These issues are discussed in
Appendices B \& C. Briefly, we use the mass function of
Sheth \& Tormen (1999; ST) and the halo profiles
of Moore et al. (1999; M99).
According to this work, the density profile of a halo interpolates
between $r^{-1.5}$ at small $r$ and $r^{-3}$ at large $r$:
$$
\rho/\rho_b = {\Delta_c \over y^{3/2} (1+y^{3/2}) }; \quad (r<r_v); \quad y\equiv r/r_c,
$$
where $\rho_b$ is the background density, $r_c$ is some
core radius, and the parameter $\Delta_c$ is a characteristic density
contrast. The virial radius, $r_v$ is defined to be the
radius within which the mean density is 200 times the background.
All these parameters are calculable functions of the halo mass,
and hence of its collapse redshift, as described in Appendix $C$.
Using these assumptions, it is possible to perform a realistic
updated version of the Neyman, Scott \& Shane calculation:
evaluate the correlations of the nonlinear density field, neglecting only the
large-scale correlations in halo positions. This is done in the next section.

\ssec{Power spectra}

An equivalent approach is to calculate the power spectrum
for this model, and this is somewhat simpler in practice. 
Start by distributing point seeds throughout
the universe with number density $n$, in which case the power spectrum of the 
resulting density field is just shot noise:
$$
\Delta^2(k)={4\pi \over n}\, \left({k\over 2\pi}\right)^3.
$$
Here, we use a dimensionless notation for the power spectrum:
$\Delta^2$ is the contribution to the fractional density variance
per unit interval of $\ln k$.  In the convention of Peebles (1980), this is
$$
\Delta^2(k)\equiv{{\rm d}\sigma^2\over {\rm d}\ln k} ={V\over (2\pi)^3}
\, 4\pi \,k^3\, |\delta_k|^2
$$
($V$ being a normalization volume), and the relation to the correlation function is
$$
\xi(r)=\int \Delta^2(k)\; {dk\over k}\; {\sin kr\over kr}.
$$

\japfig{1}{1}{wk.eps}
{The window function for the halo density
profile given by Moore et al. (1999). The
profile has $\rho\propto r^{-3/2}$ inside a 
core radius $r_c$, with $\rho\propto r^{-3}$
at larger radii, truncated at the virial radius, $r_v$.
The shape of the window function is determined by
the ratio $r_v/r_c$; the plotted lines 
correspond to 
$r_v/r_c = 2$, 4, 8, 16, 32 \& 64.}

The density field for a distribution of clumps is produced by
convolution of the initial field of delta-functions, so the
power spectrum is simply modified by the squared Fourier
transform of the clump density profile:
$$
\Delta^2(k)={4\pi \over n}\, \left({k\over 2\pi}\right)^3\; |W_k|^2,
$$
where
$$
W_k= {
\int \rho(r) \, {\displaystyle \sin k r \over \displaystyle k r}\, 4\pi \, r^2\; dr
\over
\int \rho(r) \, 4\pi \, r^2\; dr
}
.
$$
As discussed above and in Appendix C, the realistic density profiles of dark-matter
haloes are assumed to obey the M99 density profile, whose shape is 
characterized by the `concentration' $r_v/r_c$ -- the ratio of virial
and core radii.
The corresponding window functions are plotted in Fig. 1.

For a practical calculation,
we should also use the fact that hierarchical models are expected to
contain a distribution of masses of clumps, as discussed in Appendix B.
If we use the notation $n(M)\, dM$ to denote the number density
of haloes in the mass
range $dM$, the effective number density in the shot noise
formula becomes
$$
{1\over n_{\rm eff}} = { \int M^2\, n(M)\, dM \over \left[\,\int M\, n(M)\, dM\,\right]^2 }.
$$
The window function also depends on mass, so the overall power spectrum is
$$
\Delta^2(k)=4\pi\, \left({k\over 2\pi}\right)^3\; { \int M^2\, |W_k(M)|^2
\, n(M)\, dM \over \left[\,\int M\, n(M)\, dM\,\right]^2 }.
$$
The normalization term $\int M\, n(M)\, dM$ just gives the total background
density, $\rho_b$, so there is only a single numerical integral to perform.

\japfig{2}{4}{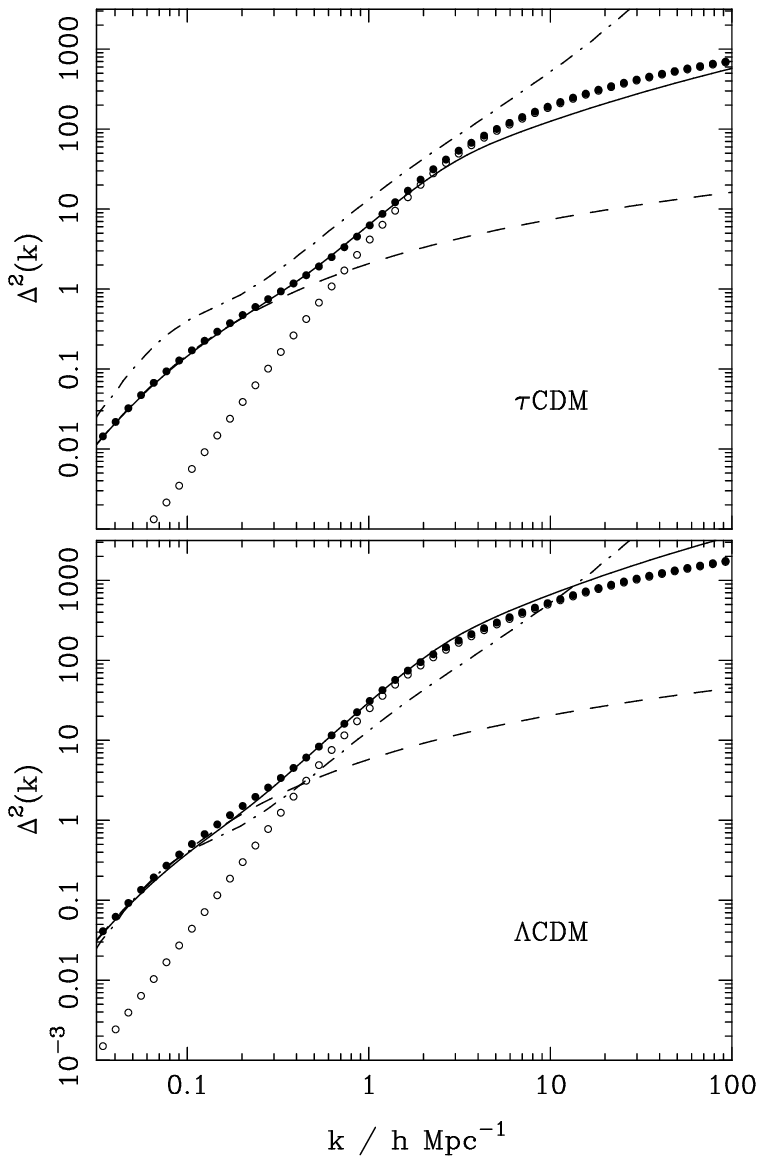}
{Density power spectrum computed for the
$\tau$CDM (top) and $\Lambda$CDM (bottom) models.
The linear spectrum is shown dashed;
the solid line shows
the nonlinear spectrum, calculated according to
the approximation of Peacock \& Dodds (1996). The spectrum according
to randomly-placed haloes is denoted by open circles;
if the linear power spectrum is added, the main
features of the nonlinear spectrum are well reproduced (filled circles).
For reference, the dot-dashed line shows the APM power spectrum 
(Maddox, Efstathiou \& Sutherland 1996).}

\ssec{CDM models}

The framework discussed so far should apply to any hierarchical
model, but the case of greatest practical interest is the family of 
CDM models. As is well known, these are characterized by a
shape parameter $\Gamma$, which is nearly but not quite $\Omega h$
when the primordial index is $n=1$ (see e.g. 
Peacock \& Dodds 1994 for a discussion of
different possible definitions for $\Gamma$). 
The normalization is specified via
the variance in fractional 
density contrast averaged over spheres of radius $R$:
$$
\sigma^2(R)=\int \Delta^2(k)\; {dk\over k}\; W_k^2,
$$
where  $W_k=3(\sin y-y\cos y)/y^3$; $y=kR$.
The abundance of rich clusters gives a measurement of 
the rms in spheres of radius $8\mpcoh$, on which
there is general agreement: 
$$
\sigma_8 = [0.5 - 0.6]\, \Omega^{-0.56}
$$
(Henry \& Arnaud 1989; White, Efstathiou \& Frenk 1993; 
Viana \& Liddle 1996;  Eke, Cole \& Frenk 1996). 
Although quite a range of these parameters remains open,
we shall focus on two commonly-discussed cases:
$\Lambda$CDM ($\Omega_m=0.3$, $\Omega_v=0.7$, $\Gamma=0.21$, 
$\sigma_8=0.9$) and $\tau$CDM ($\Omega_m=1$, $\Omega_v=0$, $\Gamma=0.21$,
$\sigma_8=0.51$). Later, we will compare with detailed simulations
of these models by Jenkins et al. (1998).

Fig. 2 shows the power spectra for these models, computed according to the above
model of randomly-placed haloes.
This turns out to agree very well with the
exact nonlinear result on small and intermediate scales.
Only for $k\ls 1 \hompc$ does the predicted power fall below
the exact result. This is only to be expected, since we have
ignored any spatial correlations in the halo positions.
A simple guess for amending this is to add  the linear power spectrum
to the power generated by the halo structure:
$$
\Delta^2_{\rm tot} = \Delta^2_{\rm random\; haloes} + \Delta^2_{\rm linear}.
$$
The justification for this is that the extra small-scale power introduced
by nonlinear evolution is associated with the internal structure of the
haloes. In practice, this model
works extremely well, giving an almost perfect description of the
power spectrum on all scales.
This is a novel way of looking at the features in the
nonlinear spectrum, particularly the steep rise between
$k\simeq 0.5\hompc$ and $k\simeq 5\hompc$, and the flattening on
smaller scales. According to the ideas presented here, the
flat small-scale spectrum arises because haloes have central
density profiles rising as $r^{-1.5}$, but not much faster.
The sharp fall in power at smaller $k$ reflects the cutoff at the virial
radii of the haloes that dominate the correlation signal.

This interpretation is quite robust and independent of the
exact form of the halo density profile used; very similar 
results are obtained using NFW haloes with central density profiles
$\rho\propto r^{-1}$ (see Appendix C), 
and even adopting the singular
isothermal density profile changes the power spectrum only a little.
This is reasonable, since a power spectrum is equivalent to
an autocorrelation: in this process, the density field is
convolved with itself, so the precise degree of cuspiness of the
central parts of the halo becomes smeared out.
This objection does not apply when we consider galaxy
correlations, however: see section 4.2.

It might be objected that this model is still not completely
realistic, since we have treated haloes as smooth objects
and ignored any substructure.
At one time, it was generally believed that 
collisionless evolution would lead to the destruction
of galaxy-scale haloes when they are absorbed into the
creation of a larger-scale nonlinear system such as a group
or cluster. However, it turns out that this
`overmerging problem' was only an artefact of
inadequate resolution (see e.g. van Kampen 2000). 
When a simulation is carried out
with $\sim 10^6$ particles in a rich cluster, the cores of
galaxy-scale haloes can still be identified after many crossing
times (Ghigna et al. 1997). This substructure must have
some effect on the correlations of the density field, and indeed
Valageas (1999) has argued that the
high-order correlations of the density field seen in
$N$-body simulations are inconsistent with a model where the
density file is composed of smooth virialized haloes.
Nevertheless, substructure seems to be unimportant at the 
level of two-point correlations.

The existence of substructure is important for the obvious next step
of this work, which is to try to understand galaxy correlations
within the current framework. It is clear that the 
galaxy-scale substructure in large dark-matter haloes defines
directly where luminous  galaxies will be found, giving hope that the 
main features of galaxy formation can be understood principally in
terms of the dark-matter distribution.
Indeed, if catalogues of these `sub-haloes' are created within a cosmological-sized
simulation, their correlation function is known to differ from that
of the mass, resembling the single power law seen in galaxies
(e.g. Klypin et al. 1999; Ma 1999).
The model of a density field consisting of smooth haloes
may therefore be an excellent description of the galaxy
field, even though it fails in detail for the dark matter as a whole.
We explore this idea in the following section.

\sec{BIASED GALAXY POPULATIONS}

In relating the distribution of galaxies to that of the mass,
there are two distinct ways in which a degree of bias is inevitable:
\smallskip
\item{(1)} Halo occupation numbers. For low-mass haloes, the
probability of obtaining an $L^*$ galaxy must fall to zero.
For haloes with mass above this lower limit, the number of
galaxies will in general not scale linearly with halo mass.

\smallskip
\item{(2)} Nonlocality. Galaxies can orbit within their
host haloes, so the probability of forming a galaxy depends
on the overall halo properties, not just the density at a point.
Also, the galaxies can occupy special places within
the haloes: for a halo containing only one galaxy, the
galaxy will clearly mark the halo centre. In general,
we will {\it assume\/} one central galaxy and a number of satellites.

\ssec{Bias parameters}

The first mechanism leads to large-scale bias, because
large-scale halo correlations depend on mass, and are some
biased multiple of the mass power spectrum:
$\Delta^2_h = b^2(M) \Delta^2$.
The linear bias parameter for a given class of 
haloes, $b(M)$, depends on the rareness of
the fluctuation and the rms of the underlying field:
$$
b=1+{\nu^2-1\over \nu\sigma}= 1+ {\nu^2-1\over \delta_c}
$$
(Kaiser 1984; Cole \& Kaiser 1989; Mo \& White 1996),
where $\nu = \delta_c/\sigma$, and $\sigma^2$ is the
fractional mass variance at the redshift of interest.
This formula is not perfectly accurate, but the 
deviations may be traced to the fact that the Press-Schechter
formula for the number density of haloes (which is assumed
in deriving the bias) is itself systematically in
error; see Sheth \& Tormen (1999) [ST], and the discussion
in Appendix B.

Note that the bias formula applies to haloes of a given $\nu$, i.e. of
a given mass. If we are interested in all haloes
{\it above\/} a given mass, we have to apply the above formula
with a weight $w_i$ for the $i$th halo:
$$
b_{\rm tot} = {\sum w_i b_i \over \sum w_i}.
$$
For a simple `censoring' -- i.e. rejecting all low-mass haloes,
but retaining all higher-mass haloes with a weight proportional
to mass, this would be
$$
b_{\rm tot} = 1 + {1\over F(>\nu)} \int_\nu^\infty {\nu^2-1 \over \delta_c}\; {dF\over d\nu}\; d\nu,
$$
Where $F(>\nu)$ is the fraction of the mass in haloes exceeding a given $\nu$;
$dF/d\nu \propto \exp(-\nu^2/2)$ according to Press-Schechter theory.
For no censoring, this gives $b=1$ exactly, as required; in general,
$b_{\rm tot}>1$.

\japfig{3}{1}{mlplot.eps}
{The empirical luminosity--mass relation
required to reconcile the observed AGS luminosity function
with two variants of CDM. $L^*$ is the characteristic
luminosity in the AGS luminosity function
(a Zwicky absolute magnitude of $-21.42$ for $h=1$).
Note the rather flat slope around
$M=10^{13}$ to $10^{14}h^{-1}M_\odot$,
especially for $\Lambda$CDM.}

Censoring also yields a bias even for Poisson-distributed
haloes. Small-scale correlations arise purely from
the correlated pairs due to the finite extent
of the haloes. Haloes of very low mass contribute no
correlated pairs except on very small scales. Thus, the
omission of the censored haloes is simply equivalent to
renormalizing the mean density, and hence scaling the correlation function.
If the fraction of particles surviving censoring is
$F(>\nu_{\rm min})$, the small-scale correlations are boosted
as follows:
$$
\xi \rightarrow \xi \; / \; [F(>\nu_{\rm min})]^2.
$$

In both cases, the natural tendency is for the galaxy distribution
to be positively biased.
The only way to achieve large-scale antibias is to give a
greater weight to the haloes with $\nu<1$ -- i.e. the efficiency
of galaxy formation has to be lower in high-mass haloes.
Small-scale antibias can be achieved via the diluting
effects of haloes whose occupation number is $N=1$. These contribute
no correlated pairs, and so simply reduce the overall correlation amplitude.
If the fraction of haloes with $N=1$ exceeds the fraction of mass
that is censored, the overall correlations will be lower than those of the mass.

\ssec{Constraints from galaxy groups}

The number of galaxies that form in a halo of a
given mass is a prime quantity that numerical models
of galaxy formation aim to calculate.
However, for a given assumed background cosmology, the
answer may be determined empirically.
Galaxy redshift surveys have been analyzed via grouping
algorithms similar to the `friends-of-friends' method
widely employed to find virialized clumps in $N$-body
simulations. With an appropriate correction for the
survey limiting magnitude, the observed number of galaxies in
a group can be converted to an estimate of the total
stellar luminosity in a group. This allows a
determination of the All Galaxy Systems (AGS)
luminosity function: the distribution of virialized
clumps of galaxies as a function of their total
luminosity, from small systems like the Local Group to
rich Abell clusters.

\japfig{4}{1}{leff.eps}
{The effective luminosity associated with galaxies
of luminosity $>L_{\rm min}$ (i.e. total luminosity
density divided by the number density of galaxies with
$L>L_{\rm min}$). A Schechter function with $\alpha=1.28$
is assumed, following Folkes et al. (1999); the
dashed line shows the effect of $\alpha=0$.
In this plot, $L^*$ is the characteristic luminosity in the
galaxy luminosity function, as distinct from the AGS $L^*$.}

The AGS function for the CfA survey was investigated by
Moore, Frenk \& White (1993), who found that the
result in blue light was well described by
$$
d\phi = \phi^*\, \left[ (L/L^*)^\beta + (L/L^*)^\gamma \right]^{-1}\;
dL/L^*,
$$
where $\phi^*=0.00126h^3\rm Mpc^{-3}$, $\beta=1.34$, $\gamma=2.89$;
the characteristic luminosity is $M^*=-21.42 + 5\log_{10}h$ in
Zwicky magnitudes. According to Efstathiou, Ellis \& Peterson (1988),
these are essentially identical to the $B_J$ magnitudes used
in the APM survey, and we assume this hereafter
(using $M_B$ to denote absolute magnitude in either of these bands).
One notable feature of this function is that it is
rather flat at low luminosities, in contrast to the
mass function of dark-matter haloes (see the discussion in Appendix B).
It is therefore clear that any fictitious galaxy catalogue
generated by randomly sampling the mass is unlikely to be a
good match to observation.
The simplest cure for this deficiency is to assume that the
stellar luminosity per virialized halo is a monotonic, but nonlinear,
function of halo mass. The required luminosity--mass
relation is then easily deduced by finding the luminosity
at which the integrated AGS density $\Phi(>L)$ matches the
integrated number density of haloes with mass $>M$.
The result is shown in Fig. 3.
The striking feature of this plot is that it is highly
non-linear:
between $M=10^{13}$ to $10^{14}h^{-1}M_\odot$,
the halo luminosity rises rather slowly, 
corresponding to a declining 
efficiency of galaxy formation over this range.
Whether or not this is physically reasonable is of
course something that can only be addressed by
detailed calculation, but
this is what the data require if CDM models are to be viable.

\japfig{5}{1}{occup.eps}
{The number density of groups as a function of the number
of galaxies they contain, for approximately volume-limited
subsamples of the CfA and ESO Slice Project group catalogues.
The dashed lines show the relation $\rho(N)\propto N^{-2.7}$.
The open points show the estimated number density of `isolated'
galaxies, i.e. those that reside in groups with $N=1$. This
was derived from the total galaxy number density, minus those
in groups (with an extrapolated contribution for $N=2$).
The solid lines show the prediction of the AGS luminosity function,
as discussed in the text.
}

\japfig{6}{1}{mass2n.eps}
{The empirical relation between halo mass and occupation number required
in order to satisfy the observed power-law distribution of group richnesses.
A limit $M_B=-19$ is assumed. 
}

\ssec{Discreteness issues}

Given a total stellar luminosity for a halo, we now need to
deduce the number of galaxies it contains. This is a
critical issue, which contains some subtle points.
The number of galaxies to be assigned to a halo of given
total luminosity can be understood easily enough for a
large halo. Suppose that the galaxy luminosity function is
of a universal form $\phi_{\ss G}$ (a Schechter function, for convenience).
If we catalogue all galaxies down to some minimum luminosity
$L_{\rm min}$, the number of galaxies found in unit volume is 
$\int_{L_{\rm min}}^\infty \phi_{\ss G} \; dL$,
whereas the total stellar luminosity is
$\int_0^\infty L\; \phi_{\ss G} \; dL$.
The number of galaxies is therefore obtained by dividing the
total luminosity by an effective luminosity per galaxy:
$$
L_{\rm eff}= { \int_0^\infty L\; \phi_{\ss G} \; dL
\over 
\int_{L_{\rm min}}^\infty \phi_{\ss G} \; dL
}.
$$
This effective luminosity is shown as a function of
galaxy sample limit in Fig. 4; clearly, from equation (19),
it depends only on the shape of the galaxy luminosity
function, and not on its normalization.

For large haloes, one would convert $M$ to $L_{\rm tot}$
as above and assign an occupation number $N=L_{\rm tot}/L_{\rm eff}$.
However, this procedure must fail for small $N$.
If we were to use integer arithmetic, the assigned occupation
number would be $N=0$ for $L_{\rm tot} < L_{\rm eff}$.
In reality, there must be a non-zero probability of finding
at least one galaxy provided $L_{\rm tot} > L_{\rm min}$.
For the low-mass haloes, the process must inevitably be stochastic,
with some haloes having $N=0$, others $N>0$.

In order to understand how to treat this situation, we must use the
observed universe as a guide.
Fig. 5 shows the distribution of galaxy groups, as a function
of the number of galaxies they contain, for two catalogues of 
groups. These are the CfA groups (Ramella, Pisani \& Geller 1997) and
the ESO Slice Project groups (Ramella et al. 1999).
The latter survey is the deeper ($B_J < 19.4$) as opposed to
the CfA limit of $B<15.5$, although the CfA sky coverage is much larger.
We have constructed approximately volume-limited subsamples
from these catalogues by considering all groups in the radial velocity ranges
$8,000 < V < 11,000 \kms$ (CfA) and $20,000 < V < 40,000 \kms$ (ESO).
Matching number densities of galaxies,
these samples correspond to approximate ($h=1$) limiting galaxy absolute magnitudes of
$M_B < -19.4$ (CfA) and $M_B < -18.5$ (ESO).
In both cases, groups are found with an algorithm similar to
friends-of-friends, so we will assume that these catalogues
approximate the distribution of occupation numbers for haloes
down to these luminosity limits. 

Results are given only for groups with $N\ge 3$, but what is striking
is that over this range the number of systems falls very fast as
a function of $N$: approximately 
$$
\rho(N)\propto N^{-2.7}.
$$
This suggests that the mean occupation number must be quite
close to unity, and this can be confirmed because the total space
density of galaxies to the above survey limits is known.
We can extrapolate the above power law to $N=2$ in order to
estimate the fraction of galaxies that exist in groups with
$N\ge 2$, and in both cases the answer is almost exactly
0.5: {\it half of all galaxies are isolated}.  This
will turn out to be important for understanding small-scale galaxy clustering:
such groups contribute no correlated pairs, and merely dilute the
overall correlations contributed by larger groups.

Can such a $\rho(N)$ dependence be predicted by the $L_{\rm eff}$
recipe discussed earlier? As expected, simple application
of integer arithmetic to the group luminosity function
counting luminosity in units of $L_{\rm eff}$ grossly underpredicts
the number of groups with $N=1$ and $N=2$ (see Fig. 5).
A detailed solution to this problem would require a knowledge of
what we may term the `conditional luminosity function', $\phi(L\mid L_{\ss G})$,
i.e. the luminosity function of those galaxies that reside in groups of
total luminosity $L_{\ss G}$ to $L_{\ss G} + dL_{\ss G}$. An observational
determination of this function is one of the results to be expected 
from future generations of redshift survey, and it can also in principle
be calculated by semianalytic galaxy formation models.
In the meantime, we can adopt an empirical approach to the problem,
based on assuming that the halo occupation number is a monotonic function
of its mass. This must be wrong in detail, but for the present heuristic
purpose it will be interesting to stick with the simplest possible
scheme. In the end, we are interested in calculating the virial radius
of the dark-matter halo in which a given galaxy group resides, and this is not a strongly-varying
function of mass; it should therefore be reasonable to ignore any scatter in
mass.

\japfig{7}{3}{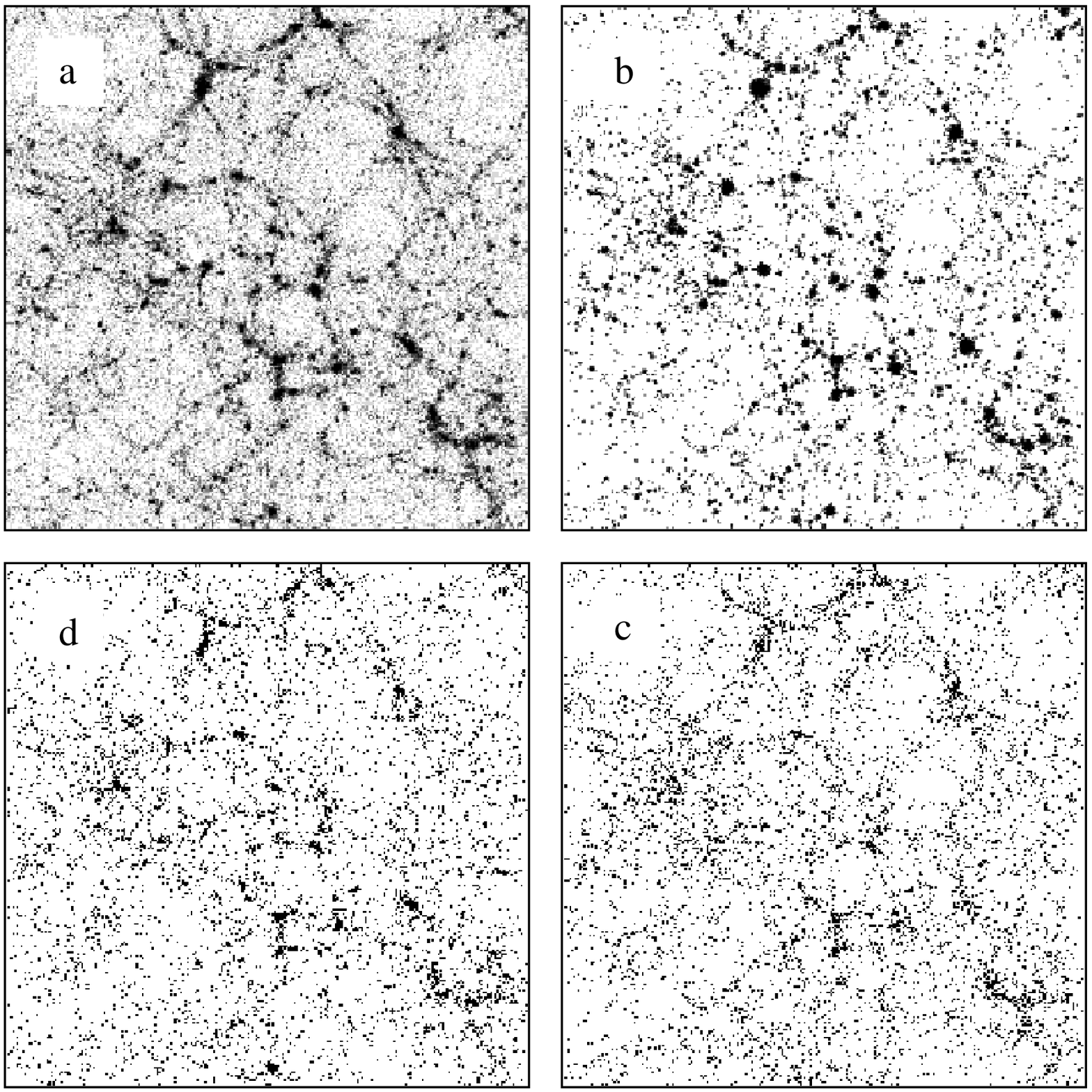}
{Stages in generating a mock galaxy sample for a
$\Lambda$CDM universe. The slices are $120\mpcoh$
on a side, with a thickness of 1/5 of the
side. Density contrast is log encoded from a value of
0.5 (white) to 10 (black). The panels show, clockwise from
top left, (a) the initial mass distribution;
(b) the mass distribution reconstructed from spherical M99
haloes, excluding the non-halo particles;
(c) the galaxy catalogue; (d) a random sampling of the mass, with the same number density
as the galaxies.}

Given this assumption, we can then assign an effective mass to a group
of given $N$ by simply calculating the integrated number density of
all groups this size or larger, $\rho(\ge N)$, and equating it to the
integrated halo number density, $n(\ge M)$. The results of this exercise
are given in Fig. 6:
$N$ is plotted as a continuous variable, but
integer arithmetic should be applied in practice to obtain the occupation
number corresponding to a halo of given mass.
To be explicit, the occupation number for a halo of given mass
is to be deduced from Fig. 6 by using the curve to deduce $N(M)$
and then taking the integer part. For example, this prescription
yields $N=1$ for a $\Lambda$CDM universe for
$10^{11.8} < M < 10^{12.6}\, h^{-1} M_\odot$.

It is interesting to contrast these results with those of
Jing, Mo \& B\"orner (1998), who were able to achieve an approximate
match between a $\Lambda$CDM model and the APM clustering, by means
of a much weaker nonlinearity in the relation between $N$
and mass: approximately $N\propto M^{0.92}$. 
However, the critical difference
is that Jing, Mo \& B\"orner (1998) did not impose a threshold in
halo mass; they allowed all simulation particles to be candidate
galaxies, even single particles that were not included in any
halo above their resolution limit. These low-mass haloes are very
weakly clustered, and dilute the correlation signal from the more
massive haloes. As a result, Jing, Mo \& B\"orner (1998) did not
need a very strong mass-dependence of the efficiency of galaxy
formation in order to achieve a reasonable clustering amplitude.
However, the need for a mass threshold seems observationally
desirable: if $L^*$ galaxies could form in 
haloes of mass $\ll 10^{12} M_\odot$, they would be baryon-dominated
and would not display flat rotation curves.
The concept of a threshold is also particularly important in
understanding voids: they must be devoid of galaxies owing to
the modification of the halo mass function in regions of
low large-scale density.
For both these reasons, we believe that our method of
predicting occupation numbers is preferable to the prescription
used by Jing, Mo \& B\"orner (1998), although they introduced
many of the correct ideas.

The occupation numbers shown in Fig. 6 will not be 
completely reliable at high masses; we have
assumed that the observed power-law distribution of group sizes
continues indefinitely, whereas there is no evidence for this
for $N\gs 20$. We therefore prefer to use the occupation
numbers predicted by the AGS analysis at large $N$.
In practice, the latter occupation numbers are slightly lower for
high-mass haloes than the ones given in Fig. 6. A convenient
approximation that matches smoothly from one to the other is
to replace $N$ from Fig. 6 by $N^{0.92}$.
This gives a recipe that can be used to generate a mock galaxy
sample from any model halo population, such that the correct group
statistics are obeyed.

\sec{APPLICATION TO NUMERICAL SIMULATIONS}

\ssec{Generation of mock galaxy catalogues}

The most direct way to implement the ideas in this paper is to 
work with the distribution of haloes found in an $N$-body
simulation. For this purpose, we have considered the data
from the Virgo consortium (Jenkins et al. 1998), concentrating
on the simulations of $\Omega=0.3$ $\Lambda$CDM and $\Omega=1$ $\tau$CDM universes in
boxes of side $239.5 \mpcoh$. Both have the same shape linear spectrum
($\Gamma=0.21$), with cluster-based normalizations of $\sigma_8=0.9$ and 0.51 respectively.
We have used a friends-of-friends code,
with a linking length of 0.2 times the interparticle separation,
to generate halo catalogues down to a minimum size of 10 particles. 
The masses of such systems
are such that they both correspond to a group luminosity of approximately
$M_B=-19.0$; it should therefore be possible to generate a mock galaxy
population down to this luminosity limit, by following the recipe given earlier.

The different stages of this process are illustrated in Fig. 7
for the case of $\Lambda$CDM.
Panel (a) shows the full mass distribution, whereas panel (b)
shows the mass distribution reconstructed from spherically-symmetric
haloes above the lower mass limit ($10^{11.8} h^{-1} M_\odot$ in this case).
We immediately see one of the main effects that contribute to bias:
there are no haloes in the voids. This censoring is a purely
gravitational effect, and is an inevitable result of the peak-background
split, in which the large-scale density modulates the mass function.
The voids will not be truly empty, but the halo mass function is 
shifted to lower masses, and the galaxy luminosity function must therefore
shift to lower luminosities. Such an effect now seems to be
clearly established observationally (e.g. Grogin \& Geller 1999).
This censoring alone would yield a positive bias, as discussed above;
this can however be offset by the nonlinear dependence of occupation number on
halo mass, although such an effect is less clear in Fig. 7.
The final galaxy catalogue (panel c) seems to the eye quite similar
to a sparse sampling of the mass (panel d), but their clustering
statistics in fact differ quite markedly, as shown below.

\japfig{8}{4}{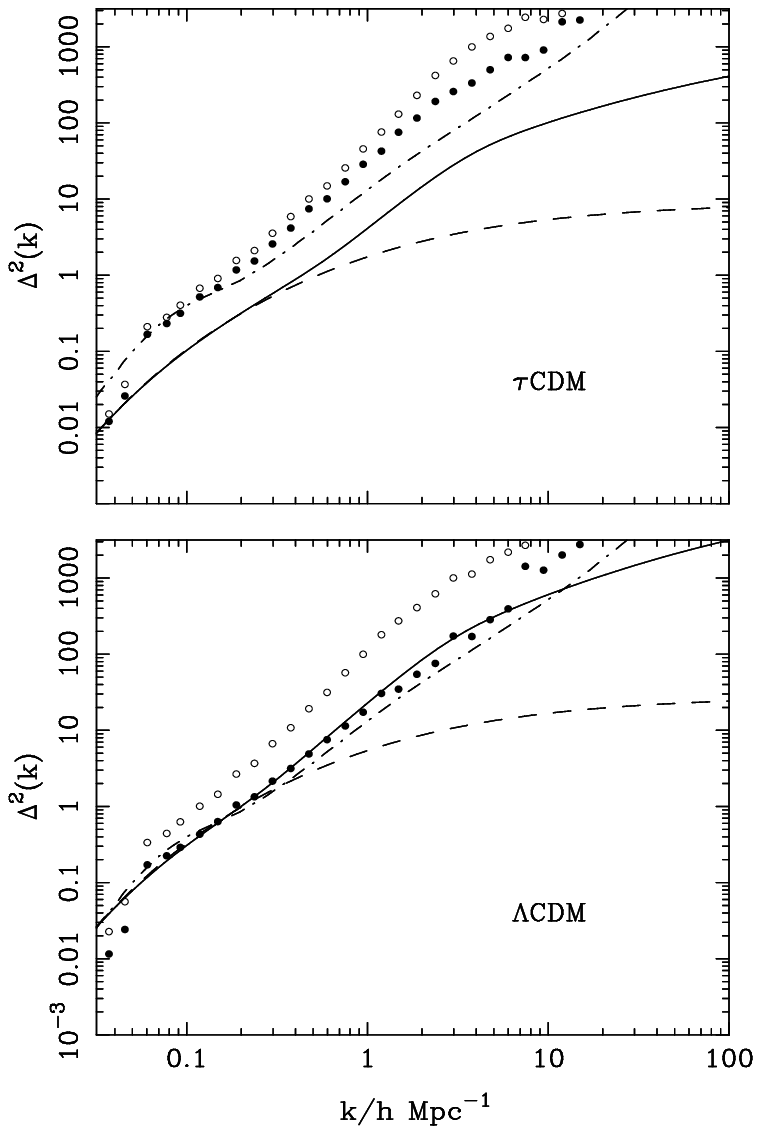}
{The power spectra for galaxy catalogues constructed from
(a) the $\tau$CDM model; (b) the $\Lambda$CDM model, using
the empirical recipe for the occupation number as a function of mass.
The model predictions are shown as filled circles;
open circles show only the effects of censoring: the power
spectrum of the mass, excluding particles in haloes below the lower 
mass limit.
The linear spectrum is shown dashed;
the solid line shows
the nonlinear spectrum, calculated according to
the approximation of Peacock \& Dodds (1996). 
The dot-dashed line shows the APM power spectrum 
(Maddox, Efstathiou \& Sutherland 1996).}

\ssec{Galaxy power spectra}

Having generated mock galaxy samples, we can now calculate their
power spectra, and see how the above simple bias prescription has
altered their clustering properties with respect to those of the mass.
Again, we shall restrict the analysis to the $\Lambda$CDM  and $\tau$CDM models.
The results are shown in Fig. 8.

We can first compute the effects of censoring -- i.e. rejecting
low-mass haloes, but otherwise giving each mass particle equal
weight. The resulting power spectra are shown in Fig. 8 
as open circles, and greatly exceed the power spectrum for all
the mass, as expected from the discussion in Section 3.1 (approximately
half of the mass is censored). Since the $\Lambda$CDM mass correlations
already exceed the APM data, this sounds like a fatal blow to that model.
However, the predicted galaxy power spectra are very much lower that
those of the censored mass, with the $\Lambda$CDM model showing the
larger shift. 
This must reflect the nonlinear $L-M$ relation
of Fig. 3; the nonlinearity is more extreme for $\Lambda$CDM than
for $\tau$CDM, and the relative contribution of high-mass
haloes is more strongly suppressed in the former model, yielding
a stronger suppression of the clustering signal.
The final predicted galaxy clustering for both models
is close to that observed in the
APM catalogue, with a shape close to a single power law.
There is a difference in amplitude, however: the  $\Lambda$CDM model shows the
required large-scale lack of bias, intermediate-scale antibias, and positive bias
for $k\gs 10 \hompc$, whereas the $\tau$CDM model shows positive bias
on all scales, especially on small scales, and exceeds the APM power
spectrum by about a factor of 2 for $k\gs 1 \hompc$.

These results make qualitative sense in terms of the above discussion.
On large scales, the different normalizations of the two models make
it inevitable that $\tau$CDM model will have significant bias,
since galaxy-scale haloes are larger-$\nu$ fluctuations in this model 
(Kauffmann, Nusser \& Steinmetz 1997).
The lack of a strong quasilinear `bulge' at intermediate wavenumbers
can be traced to the diluting effect of isolated galaxies (haloes with
occupation number $N=1$). Finally, the positive bias on the smallest
scales is a direct result of our assumption that all haloes contain
one central galaxy, with others acting as satellites.
This gives a contribution to the pair counts as a function
of radius that automatically
follows the shape of the halo density profile, so that
galaxy correlations are not expected to flatten on small scales.
The interesting point is that, for the models
shown here, these effects conspire to yield a galaxy power spectrum that
is close to the observed power-law over almost three decades of wavenumber.

In contrast to the case of mass correlations, the form
of the halo density profile at small $r$ has a critical influence
on the predicted small-scale galaxy correlations. If $\rho\propto r^{-\gamma}$
at small $r$, then a similar scaling is expected for $\xi(r)$, at
very small radii where the pair counts are dominated by pairs
between the central galaxy and the satellites.
The observed small-scale angular correlations suggest
$\gamma=1.6$ to 1.8, which is not so different from the M99
value of $\gamma=1.5$. However, if the NFW value $\gamma=1$
were to turn out to be correct, there would be no way to understand
the small-scale galaxy correlations in this model.
Inverting the argument, the observed steep small-scale $\xi(r)$
argues in favour of dark-matter haloes with rather cuspy
cores. This emphasizes the importance of independent tests of claims that
in some cases the dark matter in clusters has a
constant-density core (Tyson, Kochanski \&  dell'Antonio 1998).

The reason for the different amplitudes predicted for galaxy clustering
in the $\Lambda$CDM and $\tau$CDM
models can probably be traced to the fact that both models are
cluster normalized. This means that rich clusters with observed
$N\sim 10^2$ are forced to have equal abundances and masses.
However, the corresponding virial radius will be smaller in the
high-density model: $M\propto r_v^3 \Omega$, so that
$r_v\propto \Omega^{-1/3}$ (the core radius also obeys this
scaling, even if an $\Omega$-dependent definition of
the virial radius is adopted). 
If we ignore the slight $\Omega$
dependence of the concentration, and assume that the haloes that
dominate the correlation signal all obey this scaling, then the
quasilinear correlations will inevitably rise with $\Omega$.
For example, equation (2) suggests that the power should
scale as $\Omega^{2/3}$, which predicts a power ratio of 2.2
between $\Omega=1$ and $\Omega=0.3$.

Our conclusions thus agree to some extent with those of Benson et al. (2000a), who
found a realistic correlation function for $\Lambda$CDM, but not for $\tau$CDM.
However, their detailed conclusions regarding $\tau$CDM are completely different,
with very low clustering predicted, especially on small scales.
We believe that this is because Benson et al. were only able to match the
AGS luminosity function for their $\Lambda$CDM model; if the semianalytic
assumptions were altered so that the $\tau$CDM model also matched observed group properties,
we would then expect the predicted correlation function 
to be somewhat higher than that for $\Lambda$CDM.

In the end, our prediction for galaxy clustering in the $\Lambda$CDM
model still lies slightly above the APM data. 
One might wonder if this could be due to the luminosity
limit assumed: we have calculated results for galaxies brighter
than $M_B=-19$, whereas the APM results apply for a flux limited
sample. In principle, the model proposed here can be used to
predict how galaxy clustering varies with luminosity; however, 
for the present purposes it is sufficient to note that,
empirically, there is very little dependence in clustering
amplitude on luminosity for $M_B<-19$ (Loveday et al. 1995).
Fainter galaxies (down to $M_B=-15$) show weaker clustering,
but these will receive very small weight in a flux-limited
survey. The APM results will be dominated by galaxies around
$L^*$ ($M_B\simeq -19.7$); this is close to our
adopted luminosity limit, and we do not believe that luminosity
effects can be the cause of the mismatch in clustering amplitude.

In terms of the above
discussion, it is clear how a perfect match could be achieved:
the relative contributions of massive haloes need to be reduced
still further. We have carried out some experiments, and it appears
that $N$ for the most massive haloes $M\sim 10^{15}h^{-1} M_\odot$
would need to fall by a factor of about 2. A shift of this order is
arguably allowed by some of the uncertainties in this analysis,
since it can hardly be claimed that the $M/L$ ratios for rich
clusters are yet known to very high precision. Furthermore, 
any uncertainty in mass feeds through to the cluster $\sigma_8$
normalization of the spectrum, which will affect the predictions.
In any case, we would not wish to claim that the $\Lambda$CDM
model studied here matches the true universe precisely, but
there is gathering evidence that it may be reasonably close
(e.g. Wang et al. 2000). We therefore consider it important
to have achieved some understanding of how the necessary
scale-dependent bias can arise.

The results of this section are both good and bad news: it is satisfying to have
some understanding of why galaxy correlations behave as they do,
but it means that the clustering properties of galaxies on scales where
they are easily measured may be of restricted use in testing cosmological
models -- i.e. the amplitude of the correlations is mainly sensitive to
$\Omega$, rather than to the detailed shape of 
the underlying mass power spectrum. 
This emphasizes the importance of measurements on scales that
are well into the linear regime, so that 
something close to linear bias can be applied.

\ssec{Peculiar velocities}

Moving beyond correlation functions, the chief
longstanding puzzle concerning the galaxy distribution has concerned
the dynamical properties
of galaxies, in particular the pairwise peculiar velocity dispersion. 
This statistic has been the subject of debate, and preferred
values have crept up in recent years, to perhaps  $500 \kms$
at projected separations around 1~Mpc (e.g. Jing, Mo \& B\"orner 1998).
The predicted amplitude of peculiar velocities depends on the
normalization of the fluctuation spectrum; if this
is set from the abundance of rich clusters, then Jenkins et al. (1998)
found that reasonable values were predicted for large-scale
streaming velocities, independent of $\Omega$. However,
Jenkins et al. also found a robust prediction for the pairwise peculiar velocity dispersion
around 1~Mpc of about $800\kms$. The observed galaxy velocity
field appears to have a higher `cosmic Mach number' than the
predicted dark-matter distribution (Ostriker \& Suto 1990).

\japfig{9}{1}{vpplot.eps}
{The line-of-sight pairwise velocity dispersion for the $\Lambda$CDM model,
plotted against projected pair separation..
The top curve shows the results for all the mass (stars); the lower pair
of curves shows the predicted galaxy results, with (filled circles) 
and without (open circles)
assuming that one galaxy occupies the halo centre.
Assuming one central galaxy lowers the dispersion
significantly, but the main effect comes from the lower
efficiency of galaxy formation in high-mass haloes.
}

It is clearly of interest to see how this conclusion is
affected by the bias model adopted here. The 
line-of-sight pairwise peculiar velocity dispersion 
($\sigma_{12}$, defined in equation 16 of
Jenkins et al. 1998) is shown
in Fig. 9, where it is apparent that there is a substantial
difference between the properties of dark matter and galaxies.
The main contribution to this effect is the reduced weight given to more
massive haloes of higher velocity dispersion (see Fig. 6), although
there is also a significant contribution from the assumption
that there is one central galaxy, which thus does not gain
any peculiar velocity from the velocity dispersion of its parent halo.
A degree of `velocity bias' is thus an inevitable result of
this model. As with the power spectrum, the final $\sigma_{12}$
around 1~Mpc is slightly high compared to the figure of around
$500\kms$ preferred by Jing, Mo \& B\"orner (1998), but the main point is that
the general effects discussed here are plausibly the cause of the low
galaxy Mach number.

Again, the results of the simple model seem to be in general
agreement with the results of more detailed semianalytic studies.
Benson et al. (2000b) discuss in considerable detail the velocity
statistics for their $\Lambda$CDM model, which matches well to the
observed galaxy correlations and pairwise peculiar velocities.
Their results contrast with those of Kauffmann et al. (1999), who
performed a semianalytic calculation for the same model, yet found
much larger pairwise velocities. Benson et al. argue convincingly
that this can be traced to the larger occupation numbers assigned
to the more massive haloes by the Kauffmann et al. calculation.
In the end, this emphasizes what has often been said:
the pairwise velocity dispersion is not a very good statistic
to use, since it is rather sensitive to the contribution of the
most massive clusters.

\section{SUMMARY AND CONCLUSIONS}

The aim of this paper has been to show that, for all the sophistication
of modern understanding of gravitational instability, many of the basic
features of the cosmological density field can actually be understood via
the model introduced nearly half a century ago by 
Neyman, Scott \& Shane (1953).
Their view was of a universe that had fragmented into nonlinear haloes,
whose internal density structure determined the observed galaxy correlations.
Today, we would modify this in three ways:
(1) the haloes have a spectrum of masses, and a corresponding variation
in density structure; (2) the haloes are clustered, owing
to the large-wavelength part of the fluctuation spectrum whose
small-scale nonlinearities generated the haloes;
(3) galaxies do not simply randomly trace the mass density in the haloes.

There has been a marked recent resurgence of interest in this basic
density-clump paradigm (e.g. Sheth \& Jain 1997; Jing, Mo \& B\"orner 1998;
Valageas 1999; Yano \& Gouda 1999; 
Seljak 2000; Ma \& Fry 2000), and some of these papers independently
propose elements of the picture suggested in this work.
This surge in activity is probably traceable to the detailed
$N$-body work on the structure of haloes in CDM universes, driven by the
ability to simulate a large dynamic range in halo masses with large
numbers of particles per halo (e.g. Navarro, Frenk \& White 1996; M99; Bullock et al. 1999;
Subramanian, Cen \& Ostriker 1999).

We have shown here that the density-clump paradigm can give a
good quantitative understanding of the nonlinear correlations
of the cosmological mass field. Indeed, to some extent
the picture extends our existing understanding. 
Much recent work on nonlinear mass correlations has adopted the
`scaling ansatz' (Hamilton et al. 1991; Peacock \& Dodds 1996) in which the
small-scale correlations are assumed to obey stable clustering.
This assumption requires that the small-scale correlations
have a slope that depends on the power-law index of the
primordial linear spectrum, but this contradicts the predictions
of the density-clump model: the halo structure is universal and
thus the small-scale correlations should be independent of $n$.
The density-clump model thus suggests that stable clustering
should not be followed in practice, and this is exactly what is
seen in our recent work on simulations with scale-free
initial conditions (Smith et al., in preparation).

Extending this model to deal with the clustering of galaxies
requires additional assumptions, but many of these extra
ingredients can be constrained empirically. Two things are required:
(1) the `occupation number' of a given halo (the number of galaxies
it contains above a given luminosity threshold); (2) the location
of these galaxies within their halo. We have argued that the mean
occupation number as a function of mass can be obtained empirically
from the observed properties of groups as a function of the number of
galaxies they contain. For the second point, we have adopted the
hypothesis that one galaxy always marks the halo centre, with
its neighbours acting as satellites that follow the halo
density profile.
Both these assumptions require modification in more
realistic models: there will be some dispersion in occupation
number about the mean for haloes of a given mass, and dynamical
friction will cause galaxies to sink together within their
common halo. These effects are of course included 
automatically in semianalytic models; the detailed 
discussion of such models in e.g. Benson et al. (2000b)
suggests to us that our simplifying assumptions do
not cause much change in the predicted galaxy correlation properties, 

In any case, the model proposed here has significant heuristic value,
as it identifies a number of potential key issues in understanding
the main features of galaxy clustering:

\smallskip
\item{(1)} The galaxy distribution is inevitably biased, because
haloes of very low mass cannot house $L^*$ galaxies. Because galaxies
may orbit within their haloes out to the virial radius, the bias
is inevitably non-local on these scales.
This is not non-local bias as envisaged by e.g. Dekel \& Rees (1987):
there are no propagating non-gravitational effects between galaxies.
However, each galaxy is sensitive to the global properties of 
the halo it inhabits.

\smallskip
\item{(2)} The small-scale correlations of galaxies are very sensitive to
the distribution of galaxies within haloes. If galaxies trace mass
within haloes, their correlations are the autocorrelation of the
halo density profile, and are rather flat. Conversely, with the ansatz
of one central galaxy plus satellites, the small-scale correlations
should follow the form of the halo density profile. According to M99, this
is $\rho \propto r^{-1.5}$;  this is probably the main explanation for
the observed small-scale power-law clustering.
The success of this model is an argument against suggestions that clusters of
galaxies may not have cuspy cores (Tyson, Kochanski \&  dell'Antonio 1998).

\smallskip
\item{(3)} The amplitude of this small-scale clustering depends
critically on the number of `isolated' galaxies: haloes whose
occupation number is unity, and which thus contribute no
small-scale correlated pairs. Empirically, most galaxies seem
to exist in such a state, so it is possible to achieve an antibiased
galaxy population on intermediate scales (although the different
small-scale slopes mean that galaxy correlations always exceed those
of the mass on sufficiently small scales).

\smallskip
\item{(4)} The required occupation numbers as a function of 
mass are constrained by observations of 
the abundances of galaxy groups as a function of luminosity.
For most popular CDM models, this requires a strongly mass-dependent
$M/L$ for haloes in the range $10^{12} M_\odot$ to
$10^{15} M_\odot$. If a detailed galaxy formation model is to
yield reasonable clustering properties, it must account for this
variation. 

\smallskip
\item{(5)}
Once this empirical constraint is satisfied, 
most models predict rather similar small-scale clustering, 
but models with $\Omega\simeq 0.3$ match the data better
than models with $\Omega=1$. This will be a general
feature of cluster-normalized models, owing to the 
smaller core radii expected for haloes of a given mass
in a high-density model.

\smallskip
\item{(6)} The density-clump model also appears to account
naturally for the low small-scale pairwise velocity dispersion of galaxies.
The main effect is the down-weighting of high-mass haloes required
in CDM models in order to achieve the observed galaxy group
luminosity function.

\smallskip
\noindent
We expect that this model will be worth exploring further. There are many
statistical properties beyond the two-point level where it will
be interesting to understand the differences between the
properties of galaxies and of the mass. The model will also
serve as a useful tool for rapidly generating mock galaxy catalogues
from $N$-body simulations. As computer technology improves, more
`exact' calculations of large-scale galaxy formation will be possible, but
these will inevitably require simplified treatment of the star-formation
process, and so will never be completely robust. We believe that
the issues outlined above will continue to be important in understanding
the results from such calculations, and how they relate to the 
real distribution of galaxies.

\section*{ACKNOWLEDGEMENTS}

We are grateful to Shaun Cole, both for the seminar that prompted
this work, and also for helpful comments on the draft paper.	
We thank Pascal de Theije for the use of his 
friends-of-friends code.
The $N$-body data used in this paper were generated by
the Virgo Consortium, using supercomputers at the
Computing Centre of the Max-Planck Society in Garching,
and at the Edinburgh Parallel Computing Centre.
RES is supported by a PPARC Research Studentship.

{

\pretolerance 10000

\section*{References}

\ref Benson A.J., Cole S., Frenk C.S., Baugh C.M., Lacey C.G., 2000a, \mn, 311, 793
\ref Benson A.J., Baugh C.M., Cole S., Frenk C.S., Lacey C.G., 2000b, \mn, in press, astro-ph/9910488
\ref Bond J.R., Cole S., Efstathiou G., Kaiser N., 1991, \apj, 379, 440
\ref Bullock J.S., Kolatt T.S., Sigad Y., Somerville R.S., Kravtsov A.V., Klypin A.A., Primack J.R., Dekel A., 1999, astro-ph/9908159 
\ref Cole S., Kaiser N., 1989, \mn, 237, 1127
\ref Cole S., Arag\'on-Salamanca A., Frenk C.S., Navarro J.F., Zepf S.E., 1994, \mn, 271, 781
\ref Coles P., 1993, \mn, 262 ,1065
\ref Dekel A., Rees M.J., 1987, Nat, 326, 455
\ref Efstathiou G., Ellis R.S., Peterson B.A., 1988, \mn, 232, 431
\ref Eke V.R., Cole S., Frenk C.S., 1996, \mn, 282, 263
\ref Folkes S., et al., 1999, \mn, 308, 459
\ref Ghigna S., Moore B., Governato F., Lake G., Quinn T., Stadel J., 1998, \mn, 300, 146
\ref Grogin N.A., Geller M.J., 1999, \aj, 118, 2561
\ref Hamilton A.J.S., Kumar P., Lu E., Matthews A., 1991, \apj, 374, L1
\ref Henry J.P., Arnaud K.A., 1991, \apj, 372, 410
\ref Jenkins A., Frenk C.S., Pearce F.R., Thomas P.A., Colberg J.M., White S.D.M., Couchman H.M.P., Peacock J.A., Efstathiou G., Nelson A.H., 1998, ApJ, 499, 20
\ref Jing Y.P., Mo H.J., B\"orner G., 1998, ApJ, 494, 1
\ref Kaiser N., 1984, ApJ, {284}, L9
\ref Kauffmann G., White S.D.M., Guiderdoni B., 1993, \mn, 264, 201
\ref Kauffmann G., Nusser A., Steinmetz M., 1997, \mn, 286, 795
\ref Kauffmann G., Colberg J.M., Diaferio A., White S.D.M., 1999, \mn, 303, 188
\ref Klypin A.,  Primak J., Holtzman J., 1996, \apj, 466, 13
\ref Klypin A., Gottl\"ober S., Kravtsov A.V., Khokhlov A.M., 1999, ApJ, 516, 530
\ref Loveday J., Maddox S.J., Efstathiou G., Peterson B.A., 1995, \apj, 442, 457
\ref Ma C.-P., 1999, \apj, 510, 32
\ref Ma C.-P., Fry J.N., 2000, \apj, 531, L87
\ref Maddox S. Efstathiou G., Sutherland W.J., 1996, MNRAS, 283, 1227
\ref Mann R.G., Peacock J.A., Heavens A.F., 1998, \mn, 293, 209 
\ref Mo H.J., White S.D.M., 1996, MNRAS, 282, 1096
\ref Moore B., Frenk C.S., White S.D.M., 1993, \mn, 261, 827
\ref Moore B., Quinn T., Governato F., Stadel J., Lake G., 1999, \mn, 310, 1147 [M99]
\ref Navarro J.F., Frenk C.S., White S.D.M., 1996, ApJ, 462, 563
\ref Neyman, Scott \& Shane 1953, ApJ, 117, 92
\ref Ostriker J.P., Suto Y., 1990, \apj, 348, 378
\ref Peacock J.A., Dodds S.J., 1994, \mn, 267, 1020
\ref Peacock J.A., Dodds S.J., 1996, \mn, 280, L19
\ref Peacock J.A., 1997, \mn, 284, 885
\ref Pearce F.R., et al., 1999, \apj, 521, L99
\ref Peebles P.J.E., 1974, A\&A, 32, 197
\ref Peebles P.J.E., 1980, {\it The Large-Scale Structure of the  Universe}, Princeton Univ. Press, Princeton, NJ
\ref Press W.H., Schechter P., 1974, {\apj}, {187}, 425
\ref Ramella M., Pisani A., Geller M.J., 1997, \aj, 113, 483
\ref Ramella M. et al., 1999, A\&A, 342, 1
\ref Seljak U., 2000, astro-ph/0001493
\ref Sheth R.K., Jain B., 1997, \mn, 285, 231
\ref Sheth R.K., Tormen G., 1999, \mn, 308, 119
\ref Somerville R.S., Primack J.R., 1999, \mn, 310, 1087
\ref Subramanian K., Cen R., Ostriker J.P., 1999, astro-ph/9909279 
\ref Tyson A., Kochanski G.P., dell'Antonio I.P., 1998, \apj, 498, L107
\ref Valageas P., 1999, A\&A, 347, 757
\ref van Kampen E., Jimenez R., Peacock J.A.,  1999, \mn, 310, 43
\ref van Kampen E., 2000, astro-ph/0002027
\ref Viana P.T., Liddle A.R., 1996, MNRAS 281, 323
\ref Wang L., Caldwell R.R., Ostriker J.P., Steinhardt P.J., 2000, \apj, 530, 17
\ref White S.D.M., Rees M., 1978, \mn, 183, 341
\ref White S.D.M., Efstathiou G., Frenk C.S., 1993, {\mn}, {262}, 1023
\ref Yano T., Gouda N., 1999, astro-ph/9906375

}

\section*{APPENDIX A: HALO CORRELATION FUNCTIONS}

The correlations are easily deduced by using statistical
isotropy (see Fig. A1): calculate the excess number of pairs separated
by a distance $r$ in the $z$ direction (chosen as some arbitrary
polar axis in a spherically-symmetric clump).
Consider a point at radius $x$ in the clump; the second
point has radius $\sqrt{x^2 + r^2 + 2 x r \mu}$, where 
$\mu=\cos\theta$ is the cosine of the angle of the first point from the
polar axis. The excess number of pairs relative to random is
now easily evaluated, and the correlation function is
$$
\xi = {1\over n} \int_{-1}^1 {d\mu \over 2}
\int_0^{x_{\rm max}} \rho(\sqrt{x^2 + r^2 + 2 x r \mu}) \; {dp\over dx}\, dx.
$$
Here, $n$ is the mean number density of particles,
which is the number density of clumps times the number of
particles per clump; $\rho$ is the number density of
particles within a clump; $dp/dx$ is the radial probability
distribution for a particle in one clump. If the
clumps have a maximum radius $R$, then it can be deduced from Fig. A1 that,
for $r<R$, $x$ is unconstrained if $\mu < -r/R$, otherwise it has the upper limit
$$
x_{\rm max} = \sqrt{R^2-r^2(1-\mu^2)} - r\mu.
$$

\beginfigure{10}
\epsfxsize=5.0cm
\centerline{\epsfbox{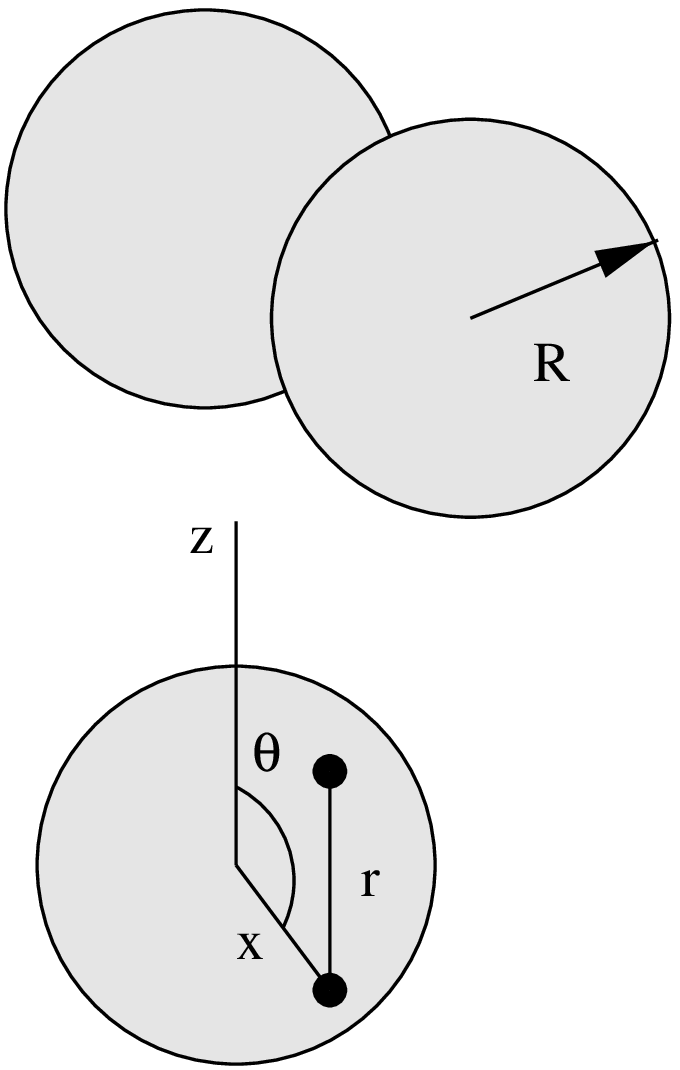}}
\caption{%
{\bf Figure A1.}
The geometry of correlations from independent haloes.
If the haloes are randomly placed (including the possibility that they
may overlap), then correlated pairs arise only within a given halo.
Consider a given point, at radius $x$ from the centre of a particular halo;
we are interested in the mean excess number of neighbours at a radius $r$ from
this point. Through isotropy, it will suffice to calculate the excess of neighbours
at an offset $r$ in the $z$ direction.
}
\endfigure

For power-law clumps, with $\rho= n B r^{-\epsilon}$, truncated at
$r=R$, the above expression for $\xi$ becomes the following for $r<R$:
$$
\eqalign{
\xi(r)=&{(3-\epsilon)B\over r^{2\epsilon-3} R^{3-\epsilon}} \, \times \cr
&\left[\; 
\int_{-1}^{-r/2R} {d\mu \over 2} 
\int_0^{R/r}
{s^{2-\epsilon}\; ds \over
(1+s^2+2s\mu)^{\epsilon/2} } \right. \; +
\cr
&\left. \int_{-r/2R}^1 {d\mu \over 2} 
\int_0^{\sqrt{\mu^2-1+R^2/r^2} -\mu}
{s^{2-\epsilon}\; ds \over
(1+s^2+2s\mu)^{\epsilon/2} } \; \right]
\cr
}
$$
(see Peebles 1974). Similar expressions apply for
$R < r < 2R$, and the correlation function vanishes for
$r>2R$.
In the limit $r\ll R$, this shows that $\xi\propto r^{3-2\epsilon}$,
provided $3/2 < \epsilon <3$, so that the $s$ integral converges at 
both small and large $s$. Values $\epsilon >3$ are unphysical, and
require a small-scale cutoff to the profile. There is no such objection
to $\epsilon < 3/2$, and the expression for $\xi$ tends to a
constant for small $r$ in this case (see Yano \& Gouda 1999).

In the isothermal $\epsilon=2$ case, Peebles (1974) showed that the integral
can be evaluated for $r\ll R$, so that the small-scale correlations in this limit
become
$$
\xi(r)={\pi^2 B\over 4 rR} = {\pi N \over 16 r R^2 n},
$$
where $N$ is the total number of particles per clump.

\section*{APPENDIX B: HALO MASS FUNCTIONS}

In recent years, it has been common practice to model the
halo mass function via the Press-Schechter (1974; PS) form:
$$
\eqalign{
f(\nu) &= \sqrt{{2\over \pi}} \, \exp[-\nu^2/2] \; \quad  \nu = \delta_c/ \sigma(R) \cr
\Rightarrow F(>\nu) &= 1-\erf(\nu/\sqrt{2}). \cr
}
$$
This gives the differential and integral fraction of the mass in the universe that 
has collapsed into objects with a mass $M$:
$$
\nu \equiv {\delta_c \over \sigma(M)},
$$  
where $\sigma(M)$ is the rms fractional density contrast
obtained by filtering the linear-theory density field on the 
required scale. In practice, this filtering is usually performed
with a spherical `top hat' filter 
of radius $R$,  with a corresponding mass of $4 \pi \rho_b R^3/3 $,
where $\rho_b$ is the background density. 
The number $\delta_c$
is the linear-theory critical overdensity, which for a `top-hat'
overdensity undergoing spherical collapse is $1.686$ -- virtually
independent of $\Omega$. 
The Press-Schechter collapsed fraction can  be converted to a differential
number density of objects, $n(M)$, using
$$
M\, n(M) = \rho_b\; {dF\over dM}.
$$

More recently, evidence has accumulated of deviations from the PS form;
Sheth \& Tormen (1999; ST) suggest the following modification:
$$
\eqalign{
f(\nu) &= 0.21617\,[ 1 + (\sqrt{2}/\nu^2)^{0.3} ] \exp[-\nu^2/(2\sqrt{2})] \cr
\Rightarrow F(>\nu) &= 0.32218\,[1-\erf(\nu/2^{3/4})] \cr
& + 0.14765\, \Gamma\,[0.2, \nu^2/(2\sqrt{2})], \cr
}
$$
where $\Gamma$ is the incomplete gamma function. A highly accurate approximation
for the integral distribution at $\nu<1$ is
$$
F(<\nu)\simeq {0.21617\nu + 0.59964\nu^{0.4} \over 1+0.073\nu^2}.
$$
Note that ST used the symbol $\nu$ to have a different meaning from
the usual $\nu = \delta_c/ \sigma(R)$, adopted here.	
Their modification partly amounts to reducing $\delta_c$
slightly, but their mass function also has a somewhat
steeper low-mass tail than the Press-Schechter formula.

\section*{APPENDIX C: HALO DENSITY PROFILES}

This appendix gives some details of the alternatives that have
been used to model the density profiles of virialized haloes.
Traditionally, virialized systems have been found by a criterion
based on percolation (`friends-of-friends'), such that the mean
density is about 200 times the mean. Sometimes, the criterion is taken
as a density of 200 times the critical value. We shall use the
former definition:
$$
r_v = \left( {3M\over 800 \pi \rho_b} \right)^{1/3}.
$$
Thus $r_v$ is related to the Lagrangian radius containing the mass via
$r_v=R/200^{1/3}$.
Of course, the density contrast used to define the
boundary of an object is somewhat arbitrary.
Fortunately, much of the mass resides at smaller radii,
near a `core radius'. These core radii are relatively
insensitive to the exact definition of virial radius.

\beginfigure{11}
\epsfxsize=8.2cm
\centerline{\epsfbox[60 208 510 588]{nfw++.eps}}
\caption{%
{\bf Figure C1.}
A comparison of various possible density
profiles for virialized haloes. The dotted line is a
singular isothermal sphere. The solid lines show
haloes with formation redshifts of 0 and 5
according to NFW ($\Omega=1$) and M99.}
\endfigure

The simplest model for the density structure of the
virialized system is the singular isothermal sphere:
$\rho= \sigma_v^2 /(2\pi G r^2)$, or
$$
\rho/\rho_b = { 200 \over 3 y^2 }; \quad (y<1); \quad y\equiv r/r_v.
$$
A more realistic alternative is the profile proposed by
Navarro, Frenk \& White (1996; NFW):
$$
\rho/\rho_b = {\Delta_c \over y (1+y)^2 }; \quad (r<r_v); \quad y\equiv r/r_c.
$$
The parameter $\Delta_c$ is related to the core radius and the virial
radius via
$$
\Delta_c = {200 c^3/3 \over \ln(1+c) -c/(1+c)}; \quad c\equiv r_v/r_c
$$
(we change symbol from NFW's $\delta_c$ to avoid confusion with the
linear-theory density threshold for collapse, and also because our definition
of density is relative to the mean, rather than the critical density).
NFW showed that $\Delta_c$ is related to collapse redshift via
$$
\Delta_c \simeq 3000 (1+z_c)^3,
$$
An advantage of the definition of virial radius used here is that
$\Delta_c$ is independent of $\Omega$ (for given $z_c$), whereas
NFW's $\delta_c$ is $\propto \Omega$.

The above equations determine the concentration,
$c=r_v/r_c$ implicitly, hence in principle giving $r_c$ in terms of
$r_v$ once $\Delta_c$ is known. A useful approximate formula for the inversion is
$$
c^{-1} \simeq 400 / (3 \Delta_c ) + (110/\Delta_c)^{0.387}.
$$
NFW give a procedure for determining $z_c$. A simplified argument would
suggest a typical formation era determined by $D(z_c)=1/\nu$, 
where $D$ is the linear-theory growth factor between
$z=z_c$ and the present, and $\nu$ is the dimensionless fluctuation
amplitude corresponding to the system in units of the rms: $\nu\equiv \delta_c/\sigma(M)$,
where $\delta_c\simeq 1.686$. For very massive systems with $\nu\gg 1$, only
rare fluctuations have collapsed by the present, so $z_c$ is close to zero.
This suggests the interpolation formula
$$
D(z_c) = 1 + 1/\nu;
$$
The NFW formula is actually of this form, except that the $1/\nu$ term
is multiplied by a spectrum-dependent coefficient of order unity.

Recently, it has been claimed by Moore et al. (1999; M99) that the
NFW density profile is in error at small $r$. M99 proposed the
alternative form
$$
\rho/\rho_b = {\Delta_c \over y^{3/2} (1+y^{3/2}) }; \quad (r<r_v); \quad y\equiv r/r_c.
$$
It is straightforward to use this in place of the NFW profile: we want to
use the same mass (and hence the same virial radius), and to
arrange for the density profiles to match at large $r$
(i.e. at the virial radius). An accurate approximation that relates the
`concentration' parameters ($c=r_v/r_c$) in the two profiles is
$$
c\,[{\rm M99}] = (c\,[{\rm NFW}]/1.7)^{0.9}.
$$
The procedure to use is therefore:

\item{(1)} pick a mass, and hence virial radius.
\item{(2)} evaluate $\nu(M)$, and hence $z_c$. 
\item{(3)} from $z_c$, get $ \Delta_c[{\rm NFW}]$ and invert to get $c[{\rm NFW}]$
\item{(4)} convert to $c[{\rm M99}]$ 
\item{(5)} from the definition of virial radius, get $ \Delta_c[{\rm M99}]$:

$$
\Delta_c \,{\rm [M99]}= {100 c^3 \over \ln(1+c^{3/2}) }.
$$

Lastly, note that the M99 profile has the practical advantage that its integral
mass distribution is readily inverted:
$$
M(<r) \propto \ln (1+y^{3/2}); \quad y\equiv r/r_c,
$$
so that it is simple to convert between $M(<r)$ and $r$ 
in either direction;
for the NFW profile, this task must be done numerically.
It may thus be more convenient to work entirely with the M99 profile, without
using NFW as an intermediate step. This requires a relation 
between $c\,{\rm [M99]}$ and $z_c$, and the following simple approximation
is accurate to a few per cent for $z_c \ls 300$:
$$
c \,{\rm [M99]} = 1.8 + 2.1 z_c.
$$

\bye